\documentclass{emulateapj}

\usepackage{xspace}
\usepackage{graphicx}
\usepackage{natbib}
\usepackage{apjfonts}
\newcommand\zth{$0^\mathrm{th}$}
\newcommand\fst{$1^\mathrm{st}$}

\newcommand{\darcsec}{\arcsec\hspace{-3pt}}
\newcommand{\darcmin}{\arcmin\hspace{-3pt}}

\newcommand\acisi{ACIS-I\xspace}

\newcommand\chandra{\textsl{Chandra}\xspace}

\newcommand\heg{{HEG}\xspace}
\newcommand\hetg{{HETG}\xspace}
\newcommand\hetgs{{HETGS}\xspace}

\newcommand\integral{\textsl{INTEGRAL}\xspace}
\newcommand\isis{\texttt{ ISIS}\xspace}

\newcommand\meg{{MEG}\xspace}

\newcommand\mysou{IGR~J18179$-$1621\xspace}
\newcommand\osa{\texttt{OSA}\xspace}

\newcommand\swift{\textsl{Swift}\xspace}

\newcommand\aproxgt{\mathrel{%
     \rlap{\raise 0.511ex \hbox{$>$}}{\lower 0.511ex \hbox{$\sim$}}}}
\newcommand\aproxlt{\mathrel{%
     \rlap{\raise 0.511ex \hbox{$<$}}{\lower 0.511ex \hbox{$\sim$}}}}

\slugcomment{Submitted }

\shorttitle{\mysou}
\shortauthors{Nowak et al. 2012}

\begin{document}

\title{X-ray and Near Infrared observations of the obscured accreting
  pulsar \mysou}

\author{M.~A. Nowak\altaffilmark{1}, 
A. Paizis\altaffilmark{2}, 
J. Rodriguez\altaffilmark{3},
S. Chaty\altaffilmark{3,4},
M. Del Santo\altaffilmark{5},
V. Grinberg\altaffilmark{6},
J. Wilms\altaffilmark{6}, 
P. Ubertini\altaffilmark{5},
R. Chini\altaffilmark{7,8}
}

\altaffiltext{1}{Massachusetts Institute of Technology, Kavli
  Institute for Astrophysics, Cambridge, MA 02139, USA;
  mnowak@space.mit.edu}
\altaffiltext{2}{Istituto Nazionale di Astrofisica, INAF-IASF, Via
  Bassini 15, 20133 Milano, Italy; ada@iasf-milano.inaf.it}
\altaffiltext{3}{AIM - Astrophysique, Instrumentation et
  Mod\'elisation (UMR-E 9005 CEA/DSM-CNRS-Universit\'e Paris Diderot)
  Irfu/Service d'Astrophysique, Centre de Saclay FR-91191
  Gif-sur-Yvette Cedex, France}
\altaffiltext{4}{Institut Universitaire de France, 103, bd Saint-Michel, 75005 Paris}
\altaffiltext{5}{Istituto Nazionale di Astrofisica, INAF-IAPS, Via
  Fosso del Cavaliere 100, 00133 Rome, Italy}
\altaffiltext{6}{Dr.~Karl Remeis-Sternwarte and Erlangen Centre for
  Astroparticle Physics, Universit\"at Erlangen-N\"urnberg,
  Sternwartstr.~7, 96049 Bamberg, Germany}
\altaffiltext{7}{Astronomisches Institut, Ruhr-Universit\"at Bochum,
  Universit\"atsstra\ss e 150, D-44780 Bochum, Germany}
\altaffiltext{8}{Instituto de Astronom\'ia, Universidad Cat\'olica del
  Norte, Avenida Angamos 0610, Casilla 1280, Antofagasta, Chile}

\begin{abstract}
  \mysou is an obscured accreting X-ray pulsar discovered by \integral
  on 2012 February 29. We report on our 20\,ksec \chandra-High Energy
  Transmission Gratings Spectrometer observation of the source
  performed on 2012 March 17, on two short contemporaneous \swift
  observations, and on our two near-infrared ($\mathrm{K}_s$,
  $\mathrm{H}_n$, and $\mathrm{J}_n$) observations performed on 2012
  March 13 and March 26. We determine the most accurate X-ray position
  of \mysou,
  $\alpha_\mathrm{J2000}=18^\mathrm{h}17^\mathrm{m}52^\mathrm{s}.18$,
  \mbox{$\delta_\mathrm{J2000}=-16^\circ 21\darcmin\, 31\darcsec.68$}
  (90\% uncertainty of 0\darcsec.6). A strong periodic variability at
  11.82\,s is clearly detected in the \chandra data, confirming the
  pulsating nature of the source, with the lightcurve softening at the
  pulse peak.  The quasi-simultaneous \chandra-\swift spectra of
  \mysou can be well fit by a heavily absorbed hard power-law
  ($N_\mathrm{H} =2.2 \pm0.3 \times 10^{23}\,\mathrm{cm}^{-2}$, and
  photon index $\Gamma = 0.4\pm0.1$) with an average absorbed
  2--8\,keV flux of $1.4\times
  10^{-11}\,\mathrm{erg}\,\mathrm{cm}^{-2}\,\mathrm{s}^{-1}$. At the
  \chandra-based position, a source is detected in our near infrared
  (NIR) maps with $\mathrm{K}_s = 13.14 \pm 0.04$\,mag, $\mathrm{H}_n
  = 16\pm0.1$\,mag, and no $\mathrm{J}_n$ band counterpart down to
  $\sim$18\,mag. The NIR source, compatible with
  2MASS~J18175218$-$1621316, shows no variability between 2012 March
  13 and March 26. Searches of the UKIDSS database show similar NIR
  flux levels at epochs six months prior to and after a 2007 February
  11 archival \chandra observation where the source's X-ray flux was
  at least 87 times fainter. In many ways \mysou is unusual: its
  combination of a several week long outburst (without evidence of
  repeated outbursts in the historical record), high absorption column
  (a large fraction of which is likely local to the system), and
  11.82\,s period does not fit neatly into existing X-ray binary
  categories.
\end{abstract}


\section{Introduction}\label{sec:intro}
\setcounter{footnote}{0}

One of the great contributions of the \integral satellite has been its
scans of the Galactic plane and bulge, which have led to the discovery
of a number of previously unknown transient X-ray binary sources
\citep{bodaghee07}. Many of these sources have high columns of
absorbing material along the line of sight, making the \integral hard
X-ray response crucial for their discovery. In numerous cases a
significant fraction of these columns are \emph{intrinsic} to the
systems in question. In fact, \integral has suggested possible
interesting relationships between absorption and binary properties: a
neutral column/orbital period anti-correlation (not unexpected as
short periods increases the likelihood that an X-ray source is more
deeply embedded within the wind/outflow from the secondary), and a
neutral column/X-ray pulsar spin period correlation \citep[which, if
real, lacks an explanation;][]{bodaghee07}.

X-ray binaries themselves represent a wide variety of types of
systems: High Mass X-ray Binaries (HMXB), which may be either wind fed
or Roche Lobe Overflow fed, Be/X-ray binaries, and accreting msec
X-ray pulsars \citep[e.g.,][]{paizis05b}, among other types of
systems. Outbursts can be very rapid, lasting only a few days in the
case of the so-called Supergiant Fast X-ray Transients (SFXT;
\citealt{negueruela06,sidoli11}, and references therein), or last
several weeks in the cases of Be/X-ray binaries and accreting msec
X-ray pulsar systems. \integral, in combination with observations at
other wavelengths, has been crucial in identifying the natures of
these sources. 

Since 2005 we have had a \chandra-followup program to observe
previously unknown X-ray sources discovered by \integral, in order to
aid in localizing and identifying these sources for multi-wavelength
follow-up observations. On 2012 February 29 (MJD 55986) \integral
discovered a new hard X-ray transient \mysou \citep{tuerler12}. The
best source position was determined with the two JEM--X instruments
with an associated uncertainty of \mbox{1\darcmin.5}. The combined
JEM--X and IBIS/ISGRI spectrum ($F_\mathrm{3-50\,keV}=1.0\times
10^{-9}\,\mathrm{erg}\,\mathrm{cm}^{-2}\,\mathrm{s}^{-1}$) could be
described by a cut-off power-law ($\Gamma=-0.5\pm0.5$,
$E_\mathrm{cutoff}= 4.9^{+1.5}_{-0.9}\,\mathrm{keV}$, where photon
flux per unit energy is $\propto
E^{-\Gamma}\exp(-E/E_\mathrm{cutoff})$) plus a broad Gaussian
absorption line ($E_\mathrm{c}={20.8}^{+1.4}_{-1.8}$\,keV).  According
to the authors, if this line is interpreted as a cyclotron resonant
scattering feature, \mysou could be a High Mass X-ray Binary (HMXB)
pulsar with a magnetic field of about $1.7 \times 10^{12}$\,Gauss. The
pulsating nature was confirmed by \swift and \textit{Fermi}/GBM
detections of a strong signal at $P=11.82\pm0.01$\,s
\citep{halpern12,li12,finger12}. The \swift observations performed on
February 29 confirmed the hard spectrum of \mysou ($\Gamma=0.4\pm0.4$)
as well as a high absorbing column density,
$N_\mathrm{H}=(11.0\pm2.0)\times10^{22}\,\mathrm{cm}^{-2}$
\citep{li12}.
 
A refined position with respect to the \integral discovery
($1\darcmin.5$ uncertainty) was reported by \cite{li12} using \swift
data. That position ($2\darcsec.2$ uncertainty) was consistent with
the position measured by \integral and was compatible with an infrared
candidate counterpart, 2MASS~J18175218$-$1621316. However, this was
not confirmed by \cite{halpern12} who reported a \swift-based source
position 4\darcsec.4 away (no error given), hence the proposed
association with the 2MASS source needed confirmation.

The day after the discovery of \mysou, we triggered our approved
\chandra target of opportunity program.  The observation was
originally set to occur on 2012 March 11, but due to a strong solar
flare it was postponed until 2012 March 17.  Thanks to the excellent
\chandra imaging and astrometry, an X-ray position with a
\mbox{0\darcsec.6} (90\% confidence level) uncertainty was reported
\citep{paizis12}. This new \chandra-based position was 0\darcsec.37
away from that obtained by \citet{li12} and 0\darcsec.09 from the
2MASS~J18175218$-$1621316 source ($\mathrm{K}_s$=13.14\,mag). The {
  preliminary} \chandra position confirmed this latter source as the
best candidate counterpart to \mysou.

\section{Observations and data analysis}
Since its discovery, \mysou has been observed by several X-ray
observatories, i.e., \integral, \swift, and \chandra. In this paper we
focus on our \chandra observations, but we include the recent
\integral long-term lightcurve in order to place the \chandra
observations within the overall context of the source
outburst. Additionally, we consider quasi-simultaneous (within a few
hours) \swift spectra. We refer to \citet{bozzo:12a} and
\citet{li:12a} for studies of the overall \swift and \integral data of
\mysou.  We also report on our near-infrared (NIR) observations of the
source field.

\subsection{\chandra data}\label{sec:chandra}

We observed \mysou for 20\,ks with \chandra from 2012 March 17,
23:09\,UT to 2012 March 18, 05:17\,UT (MJD 56003--56004, Observation ID
13684) with the High Energy Transmission Grating Spectrometer, \hetgs
\citep{canizares00}. The \hetgs creates an undispersed, CCD-spectral
resolution image (the \zth\,order), and creates high spectral
resolution, dispersed spectra in the 0.8--10\,keV band with the High
Energy Grating (\heg) and in the 0.4--8\,keV band with the Medium
Energy Grating (\meg). The \heg and \meg contain multiple spectral
orders, dispersed in opposite directions (positive and negative
orders) along the \chandra CCD detectors. Throughout this work we
shall consider data from the \zth\,order and the $\pm$\fst\, orders.
Higher spectral orders have very low count rates, and thus shall be
ignored.

Additionally, we consider a 16\,ks archival \chandra observation,
taken on 2007 February 11 (MJD 54142, Observation ID 6689), utilizing
only the Advanced CCD Imaging Spectrometer-Imaging and -Spectroscopy
arrays (\acisi and -S; i.e., the gratings were not utilized) that
contained the position of \mysou within the \acisi field of view.  We
use these data to establish an upper limit to the X-ray flux of \mysou
during an epoch prior to the outburst discussed in this work.  (There
are limited archival X-ray observations of \mysou; see
\citealt{bozzo:12a} and \citealt{li:12a} for further discussion of
\swift and \integral observations.)

Both \chandra datasets were analyzed in a standard manner (with
pixel randomization of the event positions turned off, and corrections
for Charge Transfer Inefficiency in the CCD detectors applied) using
the CIAO version 4.4 software package and \chandra CALDB version
4.4.8. The resulting spectra, which had integration times of 19.6\,ks
for the \hetgs observation and 15.8\,ks for the \acisi
  observation, were analyzed with the \isis analysis system, version
1.6.1 \citep{houck02}. Timing analysis of the data has been performed
using the \texttt{S-lang/ISIS} Timing Analysis
Routines\footnote{http://space.mit.edu/CXC/analysis/SITAR/}
(\texttt{SITAR}) package.

\subsection{\integral data}\label{sec:integral}

Starting from its discovery, \mysou has been in the \integral/IBIS
\citep{ubertini03} field of view during the \textsl{inner Galactic
  disk}, \textsl{Galactic Centre} and \textsl{Scutum/Sagittarius arms}
observations. A complete study of these \integral data is beyond the
scope of this paper; however, in order to gauge the broad-band
long-term behavior of this source we have analyzed the IBIS/ISGRI
\citep{lebrun03} and JEM-X \citep{lund03} data\footnote{We have
  analyzed available JEM-X1 and JEM-X2 data, but as they are
  consistent, we show only JEM-X1 results for clarity.}, starting from
revolution 1145 (2012 February 29, 02:20 UT, MJD 55986.1) to
revolution 1153 (2012 March 25, 21:14 UT, MJD 56011.9). A standard
analysis using version 9.0 of the Off-line Scientific Analysis (\osa)
software was performed for the pointings where the source was in the
JEM-X field of view and within $12^{\circ}$ of the center of the
IBIS/ISGRI field of view.

\begin{figure}\centering

\includegraphics[width=0.45\textwidth]{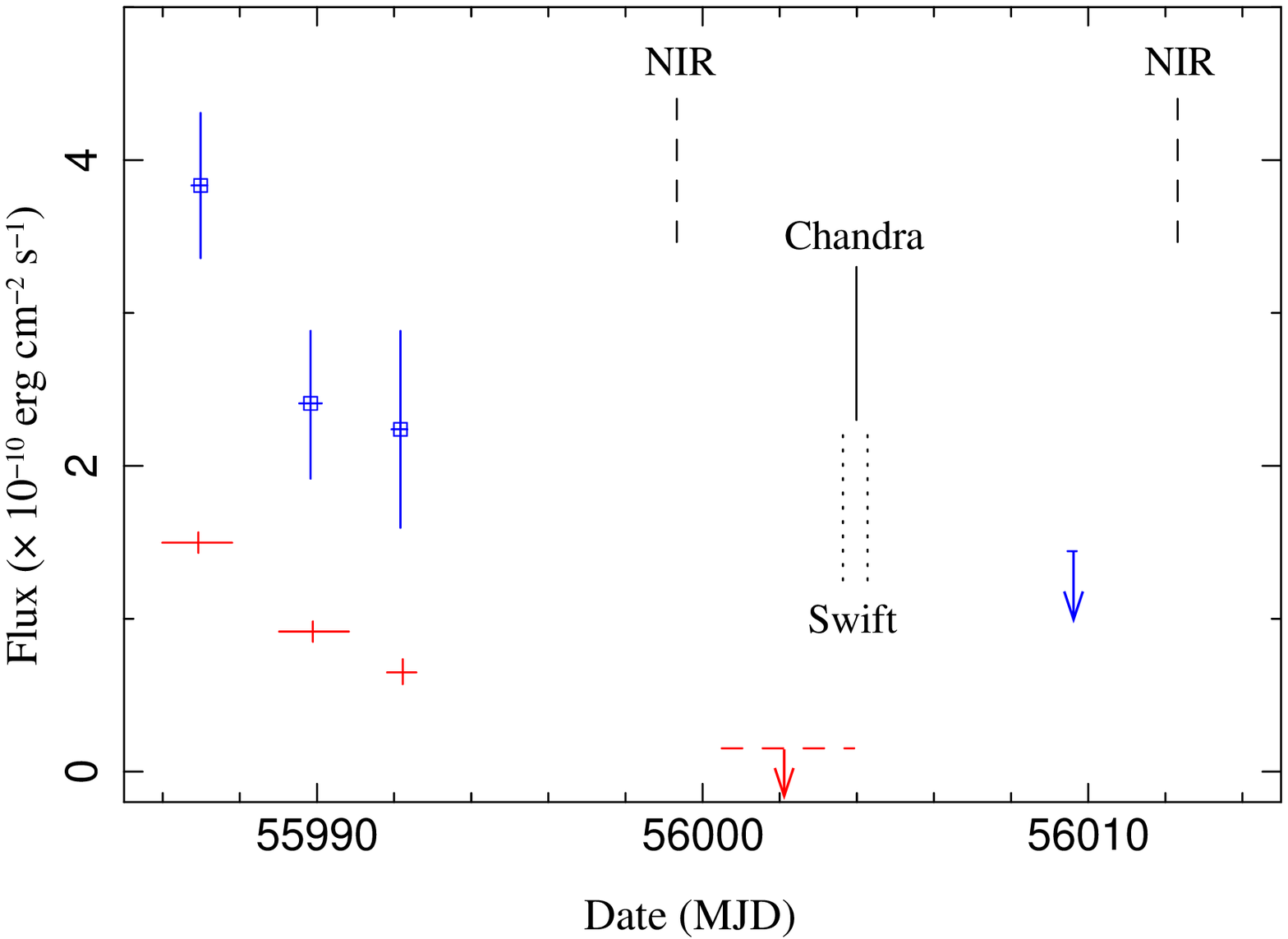}

\caption{X-ray lightcurves of \mysou as seen by \integral from its
  discovery (red crosses for IBIS/ISGRI, 18--40\,keV, and blue squares
  for JEM-X1, 3--10\,keV). Upper limits in the last observations are
  shown ($1.5\times10^{-11}\,\mathrm{erg~cm^{-2}~s^{-1}} \approx
  1.6$\,mCrab, for IBIS/ISGRI, and
  $1.5\times10^{-10}\,\mathrm{erg~cm^{-2}~s^{-1}} \approx 9$\,mCrab,
  for JEM-X1). The vertical lines show the times of the \chandra
  observation (solid line), the two quasi-simultaneous \swift
  observations (dotted lines), and the NIR observations (dashed lines)
  studied in this paper.}
 \label{fig:lcr_all}
\end{figure}

\subsection{\swift data}\label{sec:swift}

We analyzed two \swift observations performed on 2012 March 17, 15:02
UT, and 2012 March 18, 07:11 UT (obsID 00032293013 and
00032293014). The \swift/XRT data were reduced with the
\texttt{HEASOFT} suite v6.11 and the most recent calibration files
available (XRT CALDB files released on 2012 March 21). Level 2 cleaned
event files were produced from the raw data with \texttt{xrtpipeline}
with standard parameters.  The XRT was operated in photon counting
mode and the observations amount to 1301\,s and 1120\,s of good time,
respectively.  XRT source spectra were then extracted from a 20-pixel
radius circular region centered on the best source position obtained
with \chandra (see Section \ref{sec:chandrapos}).  Background spectra
were obtained from a source-free 40-pixel radius circular region.
Ancillary response files were generated for both spectra with
\texttt{xrtmkarf}, taking into account the exposure map at the source
position.  As no significant spectral variation was seen between the
two observations, the spectra were co-added and fitted as a single
spectrum together with the quasi-simultaneous \chandra spectra.

\subsection{Near-infrared  data}\label{sec:NIR}
Near-infrared $\mathrm{K}_s$, $\mathrm{H}_n$, and $\mathrm{J}_n$
observations were performed at the Universit\"atssternwarte Bochum
near Cerro Armazones in the Chilean Atacama desert. We used the 80\,cm
IRIS telescope equipped with a $1024\times1024$ pixels HAWAII-1
detector array \citep{hodapp10}. The observational sequence consisted
of eight exposures in each filter on two dates: $\mathrm{H}_n$ and
$\mathrm{K}_s$ on 2012 March 13, 08:09--08:30 UT (MJD 55999.3) and
$\mathrm{K}_s$, $\mathrm{H}_n$ and $\mathrm{J}_n$ on March 26
07:23--08:05 UT (56012.3), each comprising conventional dithering and
chopping patterns to allow subtraction of the bright NIR sky. The
total on-source integration time was 400\,s in each filter. Data
reduction involved standard IRAF procedures; astrometric and
photometric calibration were achieved via nearly 2400 and 700 sources,
respectively, from the 2MASS archive.

We also have obtained archival data from the United Kingdom Infrared
Telescope (UKIRT) Infrared Deep Sky Survey (UKIDSS; see
\citealt{lawrence:07a}, and
\citealt{casali:07a,hewett:06a,hodgkin:09a,hambly:08a} for a
description of the instruments, calibration, and pipeline).  The
source is detected during three epochs: Epoch 2006 05 01, with
$\mathrm{H}=16.45\pm0.03$\,mag; Epoch 2006 07 07, with
$\mathrm{H}=16.70\pm0.03$\,mag and $\mathrm{K}=13.007\pm0.003$\,mag;
and Epoch 2007 08 22, with $\mathrm{K}=13.043\pm0.004$\,mag.  There
are no $\mathrm{J}$-band detections of this source in the UKIDSS
database.  Below we compare these results to our recent NIR
measurements.

\section{Results}\label{sec:results}
The X-ray (3--10\,keV and 18--40\,keV) lightcurves of the 2012
outburst of \mysou as seen by \integral/JEM-X1 and \integral/IBIS,
respectively, are presented in Fig.~\ref{fig:lcr_all}. The first three
data points are the average fluxes in \integral revolutions 1145, 1146
and 1147, shown as red crosses for IBIS/ISGRI and blue squares for
JEM-X1\footnote{Revolution 1148 included \textsl{inner Galactic disk}
  observations that could not be used due to an intense solar flare
  that caused a very high radiation environment around \integral and
  forced the JEM-X and IBIS/ISGRI instruments into safe modes.}. Upper
limits in the last observations are shown: IBIS/ISGRI upper limit
($1.5\times10^{-11}\,\mathrm{erg~cm^{-2}~s^{-1}} \approx 1.6$\,mCrab
at $5\sigma$, $\sim$300\,ksec) in revolutions 1150 to 1153 and the
JEM-X1 upper limit ($1.5\times10^{-10}\,\mathrm{erg~cm^{-2}~s^{-1}}
\approx9$\,mCrab at $3\sigma$, $\sim$22\,ksec) in revolution 1153. In
all cases, the shorter exposure times of JEM-X with respect to
IBIS/ISGRI are due to the smaller field of view. We refer to
\citet{bozzo:12a} for a soft X-ray (0.3--10\,keV) \swift lightcurve of
the source outburst and for a combined \integral-\swift spectral study
of \integral revolutions 1145, 1146, and 1147. In
Fig.~\ref{fig:lcr_all}, the solid line shows the time of our 20\,ksec
\chandra observation while the two dashed lines refer to the times of
the two quasi-simultaneous \swift observations analyzed in this paper.

\subsection{\chandra}\label{sec:chandrares}
\subsubsection{\chandra position}\label{sec:chandrapos}
\begin{figure}
\epsscale{0.95} \plotone{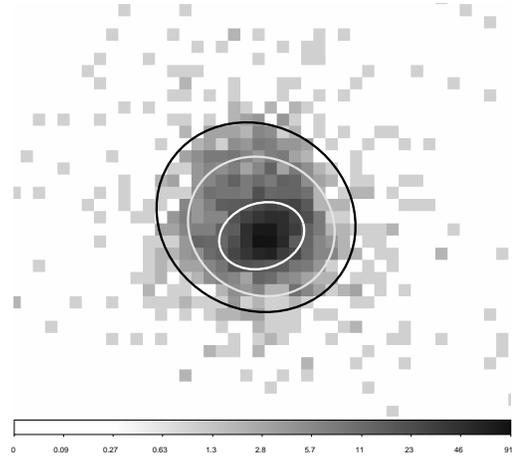}
\caption{The \chandra \zth\,order image for \mysou, using $\approx
  1/8\arcsec$ pixels. The image has been rotated so that the
  spacecraft $+Z$ coordinate points upward, and the expected \chandra
  PSF anomaly should lie in the upper left of this image. The white
  (inner) and black (outer) lines are the $2\sigma$ contours of the
  double 2D Gaussian fit to this image, while the grey (middle) line
  is the $2\sigma$ contour for the best fit single 2D gaussian.}
 \label{fig:psf}
\end{figure}

We determine the X-ray position of \mysou using the \chandra
\zth\,order image, finding
$\alpha_\mathrm{J2000}=18^\mathrm{h}17^\mathrm{m}52^\mathrm{s}.18$,
\mbox{$\delta_\mathrm{J2000}=-16^{\circ}21\arcmin 31\darcsec.68$}.
This position is 0\darcsec.143 from the one we reported in
\cite{paizis12}, which was obtained from the \texttt{tgdetect} tool
which fits a single, 2D Gaussian to the \zth\,order image. This shift
with respect to the position reported in \cite{paizis12} is due to the
fact that in this work we have more accurately taken into account a
known
artifact\footnote{{http://cxc.harvard.edu/ciao4.4/caveats/psf\_artifact.html}}
in the \chandra Point Spread Function (PSF) by fitting the \zth order
image with two asymmetric two-dimensional Gaussian functions. The
broader Gaussian function is sensitive to the expected PSF anomaly,
while the narrower Gaussian function is fit to the core of the PSF.
The asymmetry of the broad component of the PSF is seen in
Fig.~\ref{fig:psf}. This PSF anomaly is very rarely noticeable, as it
can be unveiled only when the source is bright enough to map it, but
not so bright as to distort the resulting PSF shape due to the effects
of pileup. The centroid of the narrow Gaussian function is taken as
the position of \mysou. Since the statistical errors for the
parameters of this narrow Gaussian are smaller than the absolute
pointing accuracy of \chandra, 0\darcsec.6 at 90\%
uncertainty\footnote{http://cxc.harvard.edu/cal/ASPECT/celmon/}, {
  and since there are no other point sources within this \chandra
  observation's field of view to refine further the astrometry,
  we} attribute a 90\% confidence level uncertainty of 0\darcsec.6 to
the above reported position.

\subsubsection{\chandra variability: the pulsation of \mysou}\label{sec:chandravar}

\begin{figure*}
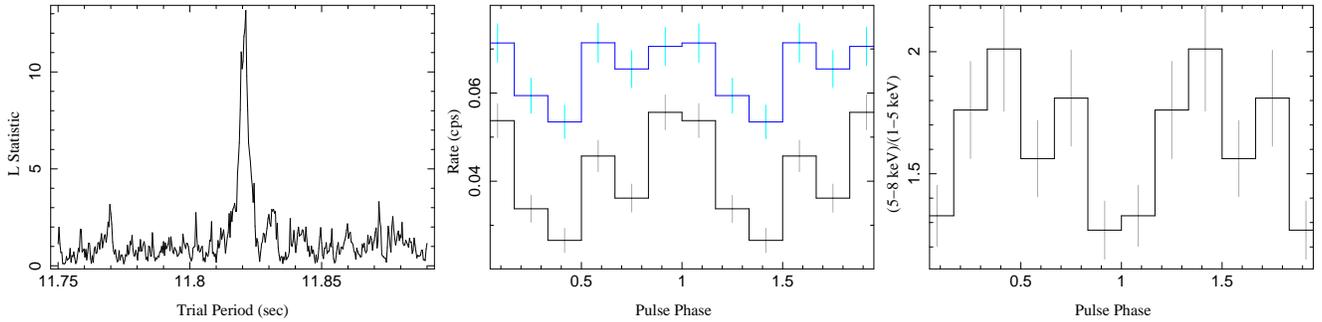
\centering
\includegraphics[width=0.32\textwidth,bb=80 15 580 379]{Epoch_Fold.ps}
\includegraphics[width=0.32\textwidth,bb=80 15 580 379]{Pulse_Lo_Hi.ps}
\includegraphics[width=0.32\textwidth,bb=80 15 580 379]{Pulse_Color.ps}

\caption{\textit{Left}: Epoch fold \protect{\citep{davies:90a}} significance
  statistic (using 6 phase bins) of the \chandra \zth\,order,
  barycenter corrected lightcurve of \mysou. The peak period is
  11.821\,s.  \textit{Middle}: The \zth\,order rate, as a function of pulse
  phase, in the 1--5\,keV band (bottom curve) and the 5--8\,keV band
  (top curve).  \textit{Right}: The ratio of the 5--8\,keV to 1--5\,keV rates
  as a function of pulse phase.}
 \label{fig:timing}
\end{figure*}

Periodic variability of the \mysou X-ray lightcurve was first reported
by \citet{halpern12} who used \swift observations to detect an
$11.82\pm0.01$\,s period and a pulsed fraction of 27\% in the
2--10\,keV band. Such a periodicity is within the realm of
detectability with our \chandra observations, which had a nominal
frame time of 1.84104\,s: a 1.8\,s exposure, followed by a 41.04\,ms
CCD readout. Actually, due to (expected and corrected on the ground)
clock drifts, the mean frame time for our observation was 1.84121\,s.
To search for the reported periodicity in \mysou, we barycentered the
lightcurve and then performed an epoch fold \citep{davies:90a} of this
lightcurve using six phase bins for the trial periods. We chose six
phase bins since, with the reported period, this gave folded
lightcurve bin durations comparable to the frame time. When folding
the lightcurve, an X-ray event is placed wholly into a single phase
bin if its barycentered time (which corresponds to the middle of the
\chandra integration frame time) falls anywhere within that phase bin.
That is, we do not add any randomization within a frame time to an
event (which would tend to smear out a signal), nor do we assign
fractional events across phase bins (which would introduce artificial
correlations among bins). Although we are working near the \chandra
temporal resolution, the fact that we sample nearly 1700 pulse periods
suggests that we should be able to determine the pulse period to an
accuracy of order 0.01\,s.

In Fig.~\ref{fig:timing} (left panel), we show the results of an epoch
fold of the \zth\,order 1--8\,keV lightcurve, extracted from a 4-pixel
radius around the point source, where we have searched 400 equally
spaced trial periods between 11.75--11.89\,s. We see that there is a
strongly detected signal at 11.82\,s, in agreement with the results of
\citet{halpern12}. The significance of this period (ignoring the
number of trials) is nominally $p=6\times10^{-6}$, compared to the
significances for the bulk of the trial values which are
$p\aproxgt0.1$. Folding the lightcurve on the most significant period
(Fig.~\ref{fig:timing}, middle panel), we see that the 1--5\,keV
X-rays show greater modulation than the 5--8\,keV lightcurve. The
standard deviation is 27\% of the mean for the former, and 11\% of the
mean for the latter. Looking at the colors of the folded lightcurve,
we see that the lightcurve softens at the peak of the pulse
(Fig.~\ref{fig:timing}, right panel). (With a mean count rate of 0.24
counts/integration frame, the \zth\,order lightcurve is slightly
affected by photon pileup, which will tend to harden the pulse peak,
not soften it. Additionally, pileup will tend to slightly decrease the
fractional amplitude of the pulse modulation.)

\subsubsection{Quasi simultaneous \swift-\chandra spectra }\label{sec:swiftchandraspe}

\begin{figure}
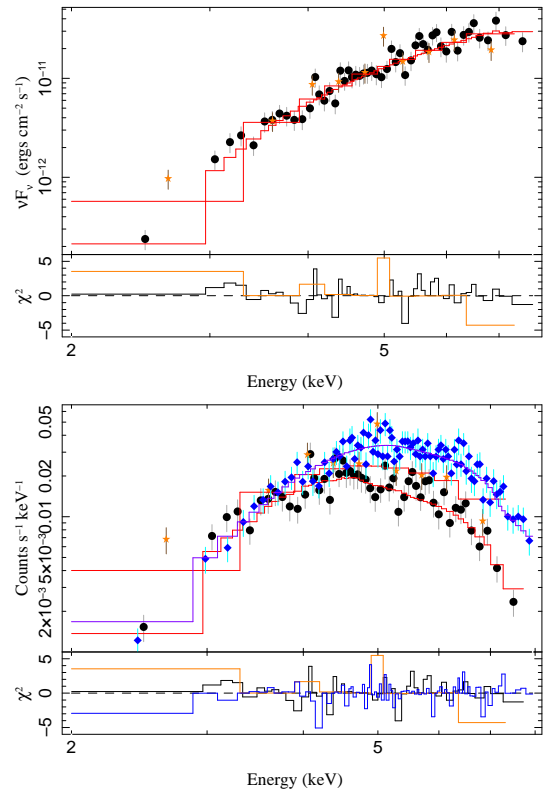
\centering

\includegraphics[width=0.4\textwidth,bb=80 15 580 379]{all_unfold.ps}
\includegraphics[width=0.4\textwidth,bb=80 15 580 379]{all_data.ps}

\caption{\textit{Top}: Flux corrected spectra for the combined
  $\pm$\fst\,order \hetg (black circles) and \swift (orange stars)
  spectra of \mysou, fit with an absorbed power-law. \textit{Bottom}:
  Detector space spectra of \mysou for the combined $\pm$\fst\,order
  \hetg, \swift, and \zth\,order (blue diamonds), fit with an absorbed
  power-law. (Additionally, a pileup model has been applied to the
  \zth\,order spectra.)}
 \label{fig:spectra}
\end{figure}

Although the folded X-ray lightcurves show evidence for a mild
spectral softening during the pulse peak, we lack the statistics to
describe this variation with spectral fits. We therefore only present
fits to the mean spectrum. We choose to fit the near contemporaneous
\swift spectra simultaneously with the \chandra spectra, and in fact
we do not even require a cross-normalization constant between these
two sets of spectra. All spectra discussed below have been binned to a
signal-to-noise ratio of 4.5 and noticed only in the 2--8\,keV band.

We find that a simple model fits the spectra well: an absorbed
power-law. We describe the absorption using the model, \texttt{TBnew},
and abundances derived from \citet{wilms00}. This model results in
higher column densities with respect to, e.g., the \texttt{wabs} model
using the cross-sections and abundances of \citet{morrison83}. Indeed,
in earlier ISM absorption models the assumed ISM abundances were
solar, while measurements outside the solar system showed that the
total gas plus dust ISM abundances are actually lower than solar
abundances. Hence with this correction, a higher column density is
needed for a given spectrum \citep[][and references therein]{wilms00}.

We further apply a pileup model \citep{davis:01a} to the \zth\,order
data, although the pileup parameter, $\alpha$, which determines the
probability of a piled event being detected as a ``good event'' is
completely undetermined within its allowed range of 0--1. This is not
unexpected as a power-law spectrum when piled appears as a slightly
harder power-law. The uncertainty in the $\alpha$ parameter merely
serves to widen the error bars on the fitted spectral slope, $\Gamma$.
 
The fitted neutral column is large, $N_\mathrm{H} = (2.2\pm0.3) \times
10^{23}\,\mathrm{cm}^{-2}$ (90\% confidence level). The fitted photon
index is extremely hard: $\Gamma = 0.4\pm0.1$ (90\% confidence level).
The fitted absorbed 2--8\,keV flux is
$(1.41\pm0.07)\times10^{-11}\,\mathrm{erg}\,\mathrm{cm}^{-2}\,\mathrm{s}^{-1}$
(90\% confidence level), i.e., $L^\mathrm{abs}_\mathrm{2-8\,keV}
\approx 10^{35}\,\mathrm{erg}\,\mathrm{s}^{-1}$ if the source is at
8\,kpc and isotropically emitting. The $\chi^2$ of the fit is 137.7
for 143 degrees of freedom. We present these fits in
Fig.~\ref{fig:spectra}. There is no strong evidence for any spectral
complexity beyond an absorbed power-law. One can add a narrow
Fe\,K$\alpha$ line to the spectra and find a 30\,eV equivalent width
(EW), but the 90\% confidence level error bars of this line encompass
0\,eV EW and are limited to $\le$71\,eV EW.

The \swift spectra agree reasonably well with the \chandra spectra,
within their limited statistics. If fit on their own, the \swift
spectra prefer a softer power-law ($\Gamma = 1.2^{+1.7}_{-1.2}$,
$\chi^2 = 8.8$ for 7 degrees of freedom); however, all spectral
parameters from the \swift-only fits have 90\% confidence level error
bars that overlap the best-fit parameters of the joint fit.
Specifically, $N_\mathrm{H} =
(2.3^{+1.9}_{-1.1})\times10^{23}\,\mathrm{cm}^{-2}$, absorbed
2--8\,keV flux =
$(1.29\pm0.20)\times10^{-11}\,\mathrm{erg}\,\mathrm{cm}^{-2}\,\mathrm{s}^{-1}$.

The above results are to be compared to the flux limits obtained from
the archival \chandra/\acisi observation of 2007 February 11.  We
extract events in a $25^{\prime\prime}$ radius region (chosen this
large to encompass the off-axis \chandra point spread function)
centered on the position of \mysou, and background events from a
nearby $115^{\prime\prime}$ radius, apparently source-free region.  In
the 2--9\,keV band there are 38 detected events, compared to an
expected 31 events from the measured background (i.e., there is a
1-$\sigma$ excess of events).  Using the same best-fit model for the
\chandra/\swift data, but leaving the model normalization free, the
90\% confidence level upper limit to the 2--8\,keV source flux during
this observation is
$1.6\times10^{-13}\,\mathrm{erg}\,\mathrm{cm}^{-2}\,\mathrm{s}^{-1}$.
Thus we see that quiescent X-ray flux levels of \mysou can be at least
87 times fainter than the flux level found with the recent
\chandra-\hetgs and \swift observations.

Searching 2.4\,Ms of IBIS/ISGRI observations and 410\,ks of JEM-X
observations of this field of view performed between 2003 February 28
and 2010 March 1, \citet{bozzo:12a} did not detect \mysou.  The
$5\sigma$ upper limits for the mean source flux over these
observations were 0.5\,mCrab ($\approx
5\times10^{-12}\,\mathrm{erg\,cm^{-2}\,s^{-1}}$) in the IBIS/ISGRI
band and 1.7\,mCrab ($\approx
2.7\times10^{-11}\,\mathrm{erg\,cm^{-2}\,s^{-1}}$) in the JEM-X band.
The latter, covering the 3--10\,keV band, is actually slightly higher
than our \chandra-\hetgs detection and significantly higher than the
90\% confidence level upper limits provided by the \acisi observation.
Still, these limits show that flux levels comparable to the beginning
of the outburst seen in Figure~\ref{fig:lcr_all} have not been typical
in \integral observations of \mysou.

\subsection{Near-infrared}\label{sec:nirres}

\begin{figure*}
\includegraphics[width=0.333\textwidth]{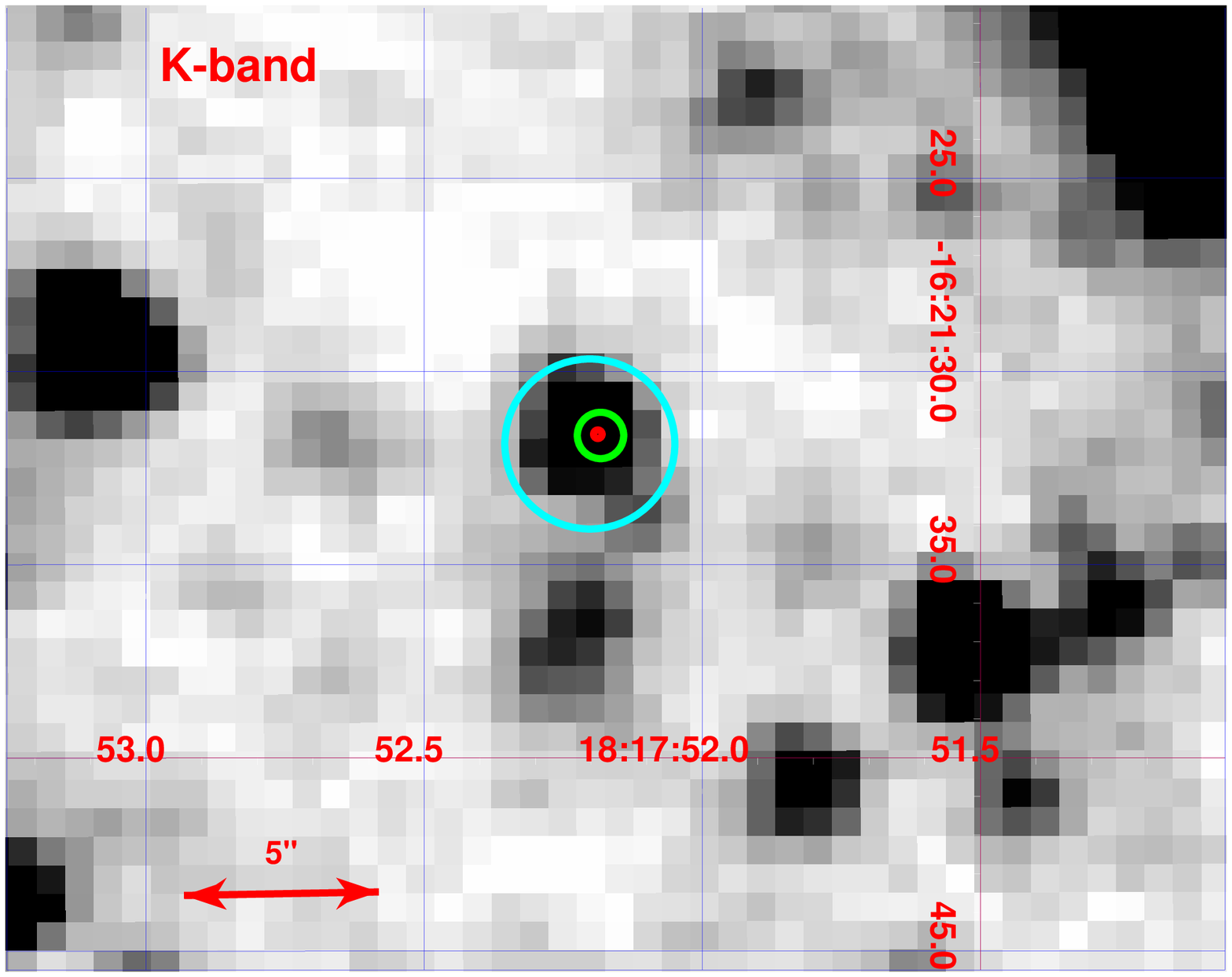}
\includegraphics[width=0.333\textwidth]{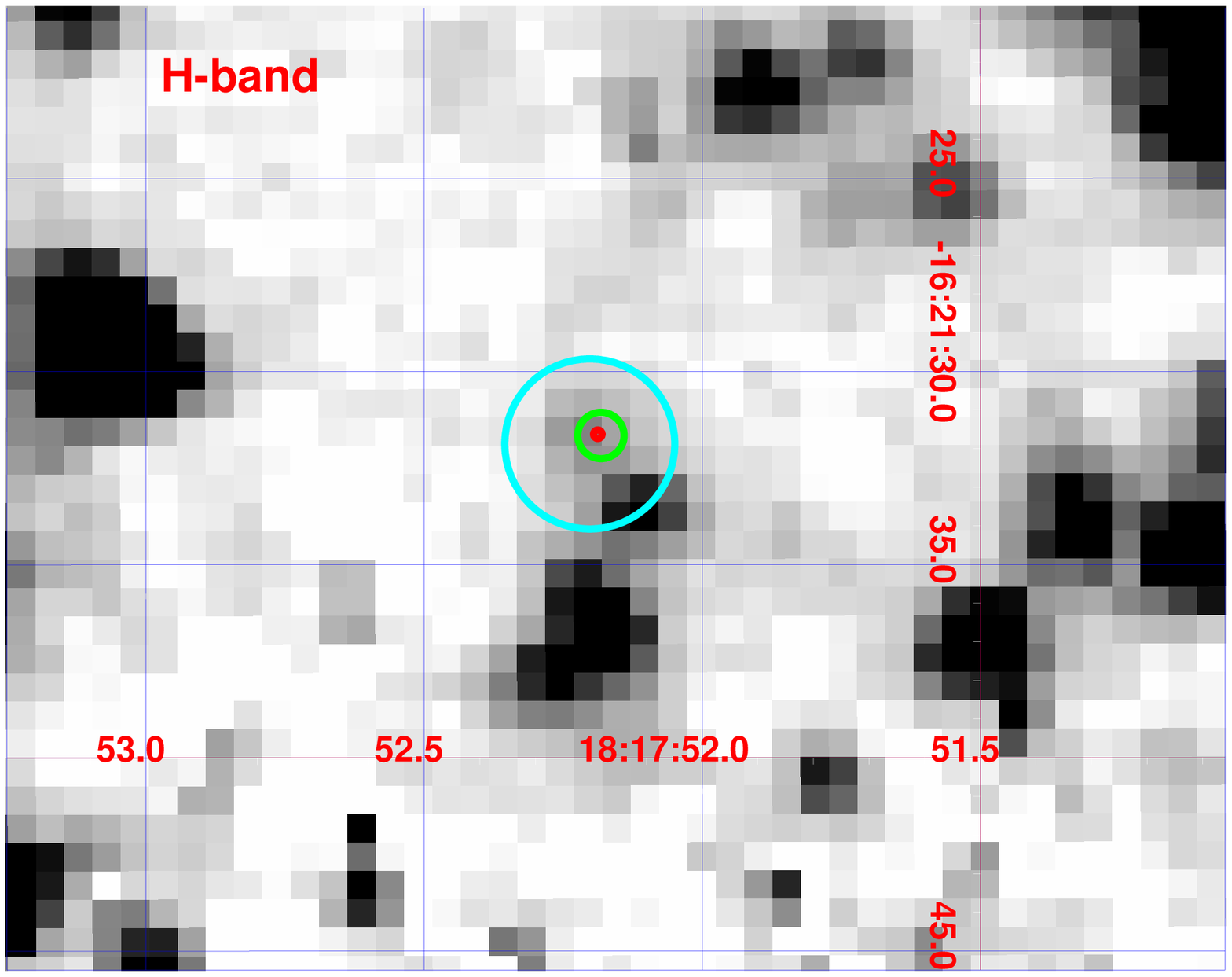}
\includegraphics[width=0.333\textwidth]{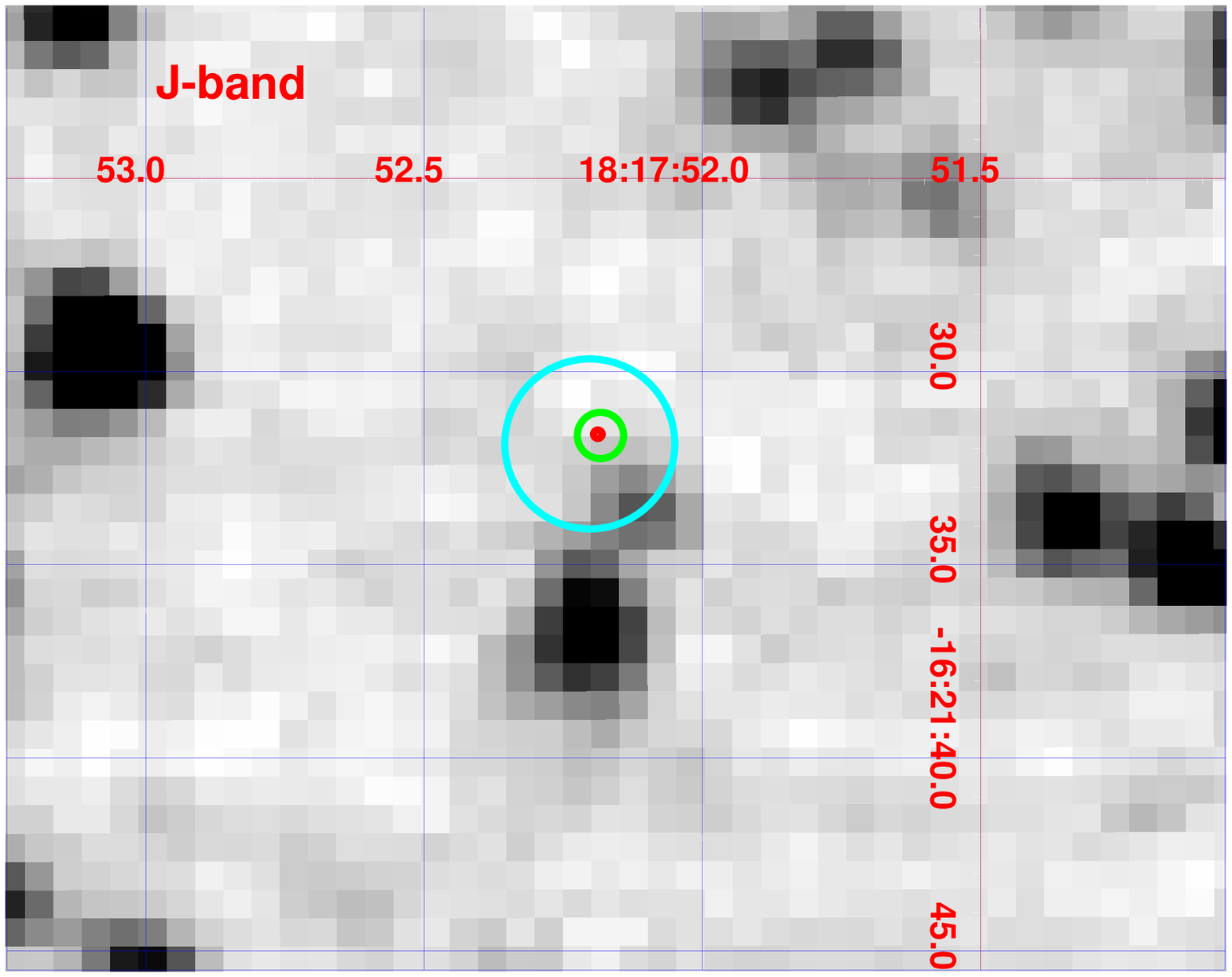}
\caption{$\mathrm{K}_s$ (\textit{left}), $\mathrm{H}_n$
  (\textit{middle}) and $\mathrm{J}_n$ (\textit{right}) band images of
  the field of view of \mysou taken on 2012 March 13. In both images
  we over-plot the \swift/XRT 90\% error circle, 2\darcsec.2 radius
  \citep[cyan,][]{li12} and the \chandra 90\% error circle,
  0\darcsec.6 radius (green, this paper). The position of the 2MASS
  J18175218$-$1621316 candidate (in red, $1\sigma$, 0\darcsec.11
  radius) lies entirely within the \chandra error
  circle. \label{fig:nir}}
\end{figure*}

The final averaged $\mathrm{K}_s$, $\mathrm{H}_n$, and $\mathrm{J}_n$
images of the field of view around \mysou, taken on 2012 March 13, are
shown in Fig.~\ref{fig:nir} and display sources down to magnitudes
$\mathrm{K}_s \sim 16$\,mag, $\mathrm{H}_n \sim 17$\,mag, and
$\mathrm{J}_n \sim 18$\,mag.

At the \chandra position of the source (0.09\darcsec\ away), there is
the already discussed 2MASS~J18175218$-$1621316 (inner most red circle
in Fig.~\ref{fig:nir}) with an observed brightness on 2012 March 13 of
$\mathrm{K}_s = 13.14 \pm 0.04$\,mag, $\mathrm{H}_n =
16\pm0.1$\,mag, and no counterpart in the $\mathrm{J}_n$ band, with
$\mathrm{J}_n <18$\,mag. The source shows no evidence of NIR
variability between 2012 March 13 and March 26.

The above observations are to be compared to the results from
UKIDSS. Although the UKIDSS database shows a 0.25\,mag variation in
$\mathrm{H}$, the IR source is quite faint and is likely contaminated
by a close by, brighter source (see Figure~\ref{fig:nir}, middle
panel). There is little variation among the $\mathrm{K}$ band
detections of UKIDSS ($\mathrm{K}=$13.007--13.043\,mag), 2MASS
($\mathrm{K}_s=$13.14\,mag), and IRIS ($\mathrm{K}_s=$13.14\,mag). The
epochs of the UKIDSS $\mathrm{K}$ band measurements bracket the
\chandra/\acisi quiescent observation, while our IRIS observation was
taken during the outburst of \mysou, yet these NIR observations are
all consistent with one another.  Also consistent with our recent
observations, UKIDSS did not detect \mysou in the $\mathrm{J}$-band.

\section{Discussion}\label{sec:discussion}

\subsection{\mysou: X-ray/NIR}\label{sec:nature}

2MASS~J18175218$-$1621316 (inner red circle in Fig.~\ref{fig:nir}) is
the only source from the 2MASS All-Sky Point Source Catalog that is
within 1\arcsec of the \chandra-determined position of \mysou, and the
2MASS point source density within 60\,$\arcsec$ of this position is
$0.01/$arcsec$^2$. Given that the \chandra error circle for \mysou is
$\approx 1$\,arcsec$^2$, the simplest hypothesis is that this 2MASS
source (with a $\sim 0.1\,\arcsec$ error radius) is the actual NIR
counterpart to the X-ray source.  There still remains, however, an
$\aproxgt 1\%$ probability that the 2MASS source is a chance
coincidence.  As we discuss below in \S\ref{sec:hmxbs}, if \mysou lies
between 1--10\,kpc distant, a wide range of companion types plausibly
would have NIR flux levels and colors of this 2MASS source, even in
the presence of large extinction.  In what follows, we therefore adopt
the hypothesis that 2MASS~J18175218$-$1621316 is indeed the NIR
counterpart to \mysou.

The 2MASS catalogue provides a source $\mathrm{K}$ magnitude of
$13.14\pm0.04$\,mag, to be compared to our constant NIR observed
$13.14 \pm 0.04$\,mag and the UKIDSS $\mathrm{K}$ magnitude of
13.007-13.043\,mag. Hence, though the X-ray outburst decay inferred
from \swift monitoring ($\sim$22\,days, \citealt{bozzo:12a,li:12a})
may resemble a low mass X-ray binary outburst type of event (accretion
disk instability), the essentially constant level of the NIR emission
(in the 2MASS and UKIDSS catalogues and in our observations during the
outburst) suggests that we are seeing the stable high mass companion
at $\mathrm{K}_s \sim 13$\,mag.  The $\mathrm{J}$-band upper limits
from our observations and UKIDSS further indicate that this companion
is highly obscured.

The heavily absorbed ($N_\mathrm{H} =(2.2 \pm 0.3) \times
10^{23}\,\mathrm{cm}^{-2}$) and hard spectrum ($\Gamma = 0.4\pm0.1$)
obtained in the joint \chandra-\swift spectral fit
(Section~\ref{sec:swiftchandraspe}), together with the 11.82\,s
pulsation (Section~\ref{sec:chandravar}) and the putative cyclotron
line discussed by \citet{bozzo:12a} and \citet{li:12a} point to an
identification of \mysou as an HMXB hosting a neutron star (NS).

The expected neutral column along this line of sight is $\approx
10^{22}\,\mathrm{cm}^{-2}$ \citep{dickey90,kalberla:05a}, which is
more than a factor of ten smaller than the measured column. A
substantial fraction of the measured neutral column is therefore
likely local to the system and associated with an
atmosphere/wind/accreting matter coming from the secondary. Associated
with this local absorption we expect an Fe\,K$\alpha$ emission line
with EW$\approx 3.3\,( {\rm N_H}/10^{22}\,{\rm cm^{-2}})$\,eV $\approx
73$\,eV, as has been found empirically in \chandra studies of HMXB
\citep{torrejon10a}. Such a Fe line strength is also expected under
the approximation that the obscuring material is spherically
distributed about the X-ray source \citep{kallman04a}. Monte Carlo
simulations of a power-law source embedded within an absorbing cloud
show the line equivalent width to be between 50\,eV and 100\,eV for
material of solar and twice solar abundances (Eikmann, priv.\ comm.).
For these reasonable assumptions, the lack of a detectable Fe
K$\alpha$ line is therefore consistent with the expectations and it
can be assumed that the bulk of the observed neutral column is local
to the system.

Figure~\ref{fig:timing} (middle and right panels) shows that the
softer 1--5\,keV folded lightcurve has greater modulation than the
harder 5--8\,keV lightcurve. That is, the lightcurve softens at the
peak of the 11.82\,s pulse.  Such an energy dependent pulse profile is
seen in many magnetized neutron stars, where the stronger variability
at lower energies is typically attributed to absorption in the
accretion column, while the (typically sinusoidal) emission pattern at
harder energies is due to emission from the bottom of the accretion
column \citep[][and references therein]{caballero:11a}.

\subsection{The Nature of \mysou: an HMXB?}\label{sec:hmxbs}

HMXBs are known to be divided into different groups according to the
nature of the companion, formation and geometry of the system
\citep[e.g.,][and references therein]{chaty11}. They include BeHMXBs
X-ray transients, usually hosting a NS on a wide eccentric orbit
around a B0--B2e spectral type donor star with a decretion disk; and
supergiant HMXBs (sgHMXBs), hosting a NS on a circular orbit around a
supergiant OB star. sgHMXBs can be either Roche lobe overflow sources
(RLO; rare transient sources with outburst luminosities of
$L_\mathrm{X}\sim10^{38}\,\mathrm{erg}\,\mathrm{s}^{-1}$), or wind-fed
systems (e.g., the majority of sgHMXB, with persistent luminosities of
$L_\mathrm{X}\sim10^{35-36}\,\mathrm{erg}\,\mathrm{s}^{-1}$).

\begin{figure}\centering
\includegraphics[width=0.4\textwidth]{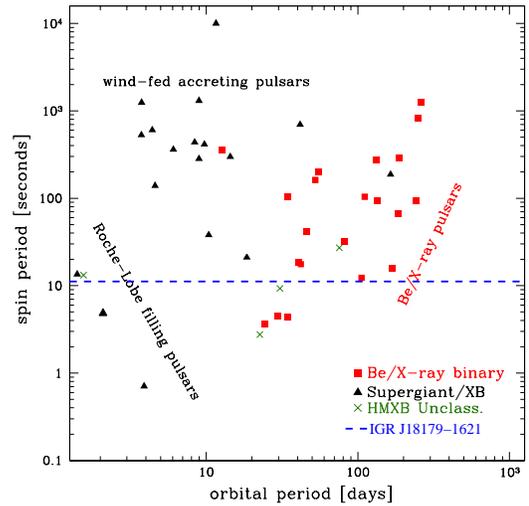}
\caption{Corbet diagram (NS $P_\mathrm{spin}$ versus system
  $P_\mathrm{orb}$; \citealt{corbet86}) showing the three HMXB
  populations: BeHMXB, sgHMXB-Roche lobe overflow and sgHMXB-wind
  fed. See text. \mysou ($P_\mathrm{orb}$ unknown) lies along the
  horizontal blue line at $P_\mathrm{spin}\sim$12\,s.  Adapted from a
  figure by J. A. Zurita Heras, as originally presented by
  \protect{\citet{chaty11}}.}
 \label{fig:corbet}
\end{figure}

An efficient way to visualize all such sources is via the so-called
Corbet diagram \citep{corbet86}, where the NS spin period,
$P_\mathrm{spin}$, is plotted versus the system orbital period,
$P_\mathrm{orb}$ (Fig.~\ref{fig:corbet}). In general, in such a
diagram BeHMXBs populate the region with higher $P_\mathrm{orb}$
(roughly $P_\mathrm{orb}> 20$\,days), while sgHMXBs populate the lower
part ($P_\mathrm{orb} < 20$\,days). $P_\mathrm{spin}$ ranges from
about 1\,s to $10^{3}$\,s for both classes of sources
(Fig.~\ref{fig:corbet}; see also \citealt{chaty11, sidoli11}). The
majority of transient sources with spin periods $\aproxlt 20$\,s
are in fact BeHMXB.  

In the last decade, two new classes of wind accreting pulsars in
sgHMXBs have been discovered \citep{bodaghee07}: obscured, fairly
persistent sources with a huge intrinsic local extinction, and sgHMXBs
exhibiting fast and transient outbursts, known as Supergiant Fast
X-ray Transients \citep[SFXT, see e.g.,][for a recent review,
especially Fig.~2 therein]{sidoli11}. The neutral column of \mysou is
typical of this latter category of sources, but its spin period -- and
prolonged outburst -- is not.

Based on the above discussion, \mysou does not clearly fall into any
of the above classes. Its transient nature seems to exclude the highly
obscured and standard wind-fed systems that are fairly persistent. A
few week long transient outburst of \mysou could suggest that it is a
BeHMXB, with its 11.82\,s spin period placing it at the lower end of
orbital periods in the Corbet diagram, i.e., several tens of days
(20--80\,days).  Although many Be X-ray binaries show periodic
outbursts, many other Be show long periods of quiescence. A good
example is A0535+26, which has had only five major outburst phases
since its 1975 discovery, which were separated by many years of
quiescence \citep[e.g., between 1994 February and 2005 May, see][and
  references therein]{caballero:07a}.

On the other hand, despite its transient nature \mysou would be
unusual for a SFXT, since these sources are known to have very fast
outbursts, on the order of hours or a few days at most, as seen by
\integral/IBIS. Fig.~\ref{fig:lcr_all} shows that \mysou is detected
for at least three revolutions, i.e., about 12 days. An extremely
nearby SFXT, however, may not be excluded.

RLO sources, i.e., transient sgHMXBs strictly filling their Roche-Lobe
(pure accretion disk accretors) are rare, since the mass transfer is
highly unstable and the accretion should only last for a few thousand
years. Instead, there are HMXBs exhibiting beginning atmospheric Roche
Lobe overflow, where the massive star does not fill its Roche Lobe,
but the stellar wind follows the Lagrange equipotentials, and
accumulates, forming an accretion disk \citep{bhattacharya1991}. This
situation is more stable, but still rare due to the required
configuration of stellar radius, orbital distance and mass ratio. We
know only three such systems in total, hosting NSs: LMC~X-4, Cen~X-3
and SMC~X-1. \mysou could be the fourth source belonging to this
class.

\begin{figure}\centering
\includegraphics[width=0.36\textwidth,angle=90]{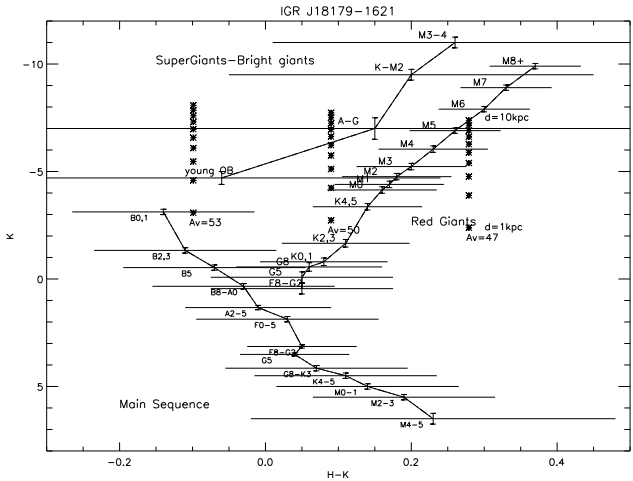}
\caption{Plot of expected H$-$K colors vs.\ absolute K magnitude for
  different stellar types. Superimposed on this are points (asterisks)
  for the NIR measurements of \mysou, assuming three different visual
  extinctions ($A_\mathrm{V}=$47, 50, and 53\,mag) for assumed
  distances ranging from 1--10\,kpc.}
 \label{fig:distance}
\end{figure}

Finally, we consider the possible stellar companions in \mysou, based
upon the NIR measurements coupled with assumptions about possible
distances and assumed extinctions. In Fig.~\ref{fig:distance}, we show
absolute K-magnitude vs.\ H$-$K colors, for a series of stellar
sequences. Overplotted on this are the absolute K-magnitude and H$-$K
colors for the counterpart of \mysou given our measurements, and
assuming distances ranging from 1--10\,kpc. 

We estimate the extinction towards the source from the relationship of
\citet{predehl95}, modified to account for the fact that absorption
model of \citet{wilms00} fits neutral columns $\approx 30\%$ larger
than the model used by \citet{predehl95}. Specifically, we assume
$A_\mathrm{V} \sim N_\mathrm{H}/2.7\times10^{21}\,\mathrm{cm}^{-2}$.
The 21\,cm column translates to an extinction of
$A_\mathrm{V}=3.7$\,mag, while the full fitted neutral column yields
$A_\mathrm{V}\sim 82$\,mag. Note, however, that the \citet{predehl95}
relationship is based on measurements of X-ray dust scattering halos.
Source intrinsic absorption in the stellar wind would happen in a
medium that does not contain dust and thus for such a case the
\citet{predehl95} relationship overestimates the $A_\mathrm{V}$.

Figure~\ref{fig:distance} shows that the range of extinctions from
$A_\mathrm{V} =47$--52\,mag is the only range where our NIR
measurements intersect standard stellar sequences. While \mysou could
be consistent with a Low Mass X-ray Binary (LMXB), with a red giant
companion ranging from K-type ($A_\mathrm{V} \sim 49$, at 1\,kpc) to
M-type ($A_\mathrm{V} \sim 47$, at 10\,kpc), a LMXB nature seems
unlikely, given the 11.82 s pulsation and very hard X-ray spectrum of
\mysou. More realistic models for \mysou are that it is a
HMXB. Supergiant companions can range from a close by OB-type
($A_\mathrm{V} \sim 52$, at 2\,kpc) to a further away A-type
($A_\mathrm{V} \sim 49$, at 7\,kpc). Even a nearby B-type main
sequence star is allowed ($A_\mathrm{V} \sim 52$), at 1\,kpc.

In all these cases, $A_\mathrm{V}$ is $\sim$50\% of what is obtained
by calculating $A_\mathrm{V}$ from the X-ray $N_\mathrm{H}$. This
result argues that about half of the observed column is caused by the
neutron star being embedded in the stellar wind of an early-type
donor, while the other half of the column would be in the (dusty)
interstellar medium close to the binary. The deduced column for the
wind absorption, $N_\mathrm{H, wind}\sim 10^{23}\,\mathrm{cm}^{-2}$,
is consistent with that seen in a number of HMXB.  For example, the
XRB GX 301$-$2, which orbits a B1 Ia hypergiant in a 41.5\,d
ellipitical orbit, has an $N_\mathrm{H}$ that is strongly variable,
with $N_\mathrm{H}\sim 10^{24}\,\mathrm{cm}^{-2}$ during the
pre-periastron flare \citep{fuerst:11a} and $N_\mathrm{H}\sim
10^{23}\,\mathrm{cm}^{-2}$ and lower after the periastron passage
\citep{suchy:12a}.  Unfortunately, we lack detailed observations over
the course of the outburst that might have revealed variations of the
neutral column.  The nature of \mysou at this time still remains
ambiguous.

\acknowledgments We thank the \chandra team for their rapid response
in scheduling and delivering the observation, as well as the \integral
Science Data Center for their quick and efficient sharing of \integral
results. AP thanks Enrico Bozzo for useful discussion and for the
support in the early phases of the source outburst that led to the
\chandra trigger. We thank the referee for comments that improved this
manuscript. MAN acknowledges the support of NASA Grants GO2-13035X and
SV3-73016. MDS acknowledges financial contribution from the agreement
ASI-INAF I/009/10/0 and from PRIN-INAF 2009 (PI: L.  Sidoli). VG and
JW acknowledge support by the Bundesministerium f\"ur Wirtschaft und
Technologie under Deutsches Zentrum f\"ur Luft- und Raumfahrt grant 50
OR 1113. This paper is partly based on observations with \integral, an
ESA project with instruments and science data center funded by ESA
member states, Czech Republic and Poland, and with the participation
of Russia and the USA. This research has made use of the \integral
sources page http://irfu.cea.fr/Sap/IGR-Sources/ and the \integral
Spiral Arm pages
http://sprg.ssl.berkeley.edu/$\sim$bodaghee/isa/. This publication
makes use of data products from the Two Micron All Sky Survey, which
is a joint project of the University of Massachusetts and the Infrared
Processing and Analysis Center/California Institute of Technology,
funded by the National Aeronautics and Space Administration and the
National Science Foundation. AP and PU acknowledge financial
contribution from the ASI-INAF agreement I/033/10/0.



\end{document}